\documentclass[a4paper,12pt]{article}
\usepackage[pctex32]{graphics}

\newcommand{\beq}{\begin{equation}}
\newcommand{\eeq}{\end{equation}}
\newcommand{\beqa}{\begin{eqnarray}}
\newcommand{\eeqa}{\end{eqnarray}}

\begin{titlepage}

\begin{document}
{\baselineskip20pt


\vskip .6cm

\begin{center}
{\Large \bf Quasinormal modes from potentials surrounding the
charged dilaton black hole }

\end{center} }

\vskip .6cm
 \centerline{\large Yun Soo Myung$^{1,a}$,
 Yong-Wan Kim $^{1,b}$,
and Young-Jai Park$^{2,c}$}

\vskip .6cm

\begin{center}
{$^{1}$Institute of Basic Science and School of Computer Aided
Science,
\\Inje University, Gimhae 621-749, Korea \\}

{$^{2}$Department of Physics and Center for Quantum Spacetime,
\\Sogang University, Seoul 121-742, Korea}
\end{center}

\vspace{5mm}


\begin{abstract}
We clarify the purely imaginary quasinormal frequencies of a
massless scalar perturbation on the 3D charged-dilaton black
holes.  This case is quite interesting because the potential-step
appears outside the event horizon similar to the case of the
electromagnetic perturbations on the large Schwarzschild-AdS black
holes. It turns out that the potential-step type provides the
purely imaginary quasinormal frequencies, while the
potential-barrier type gives the complex quasinormal modes.

\end{abstract}

\noindent PACS numbers: 04.70.Dy, 04.60.Kz, 04.30.Nk  \\
\vskip .1cm \noindent Keywords: dilaton, charged black holes,
quasinormal modes

\vskip 0.8cm

\noindent $^a$ysmyung@inje.ac.kr \\
\noindent $^b$ywkim65@gmail.com \\
\noindent $^c$yjpark@sogang.ac.kr

\noindent
\end{titlepage}

\setcounter{page}{2}

\section{Introduction}
The no-hair theorem~\cite{RW} implies that the external field of a
black hole  relaxes to a black hole spacetime characterized by
three parameters of mass, charge, and angular momentum. In other
words, the perturbations left outside the black hole would either
be radiated away to infinity or be swallowed by the black hole.
This indicates the boundary conditions in asymptotically flat
spacetimes: purely outgoing waves at infinity and purely ingoing
waves near the event horizon. These boundary conditions for
asymptotically AdS spacetimes are changed as  no waves at infinity
and purely ingoing waves near the event horizon because of the
presence of infinitely height potential at infinity. We call the
relaxation phase in the dynamics of perturbed black holes as
``quasinormal ringing"~\cite{Vish,Press}, damped harmonic
oscillations with complex frequencies. The perturbation decays and
its frequencies are complex  since the perturbation field can fall
into black hole or radiate to infinity. At late times, all
perturbations except a few small modes are radiated away in the
similar manner of the last pure dying tones of a ringing bell.

On the other hand, the  dynamics of the  Schwarzschild black hole
perturbations could be represented  by the Regge-Wheeler equation
by introducing a tortoise coordinate
$r_*$~\cite{ReggeW,Zer,Vis,Chan,Kwon}. This corresponds to the
Schr\"odinger equation, which provides the scattering problem of
the perturbation off the potential $V(r)$ surrounding the event
horizon without considering any backreaction. We have the
$T(\omega)$ transmission and $R(\omega)$ reflection amplitudes.
The greybody factor (absorption cross section) could be obtained
using the transmission amplitudes. The quasinormal modes (QNMs)
could be read off from the scattering resonances \cite{hod}.
Specifically, the pole structure of these two amplitudes reflects
that the QNMs correspond to purely ingoing modes near
the horizon and purely outgoing modes at infinity. In this sense,
the potential contains all information on the black hole
background.

There had been extensive works done to compute QNMs and to analyze
them in various black hole backgrounds~\cite{kokko}. One of the
reasons for the attention on QNMs is the conjecture relating
anti-de-Sitter space (AdS) and conformal field theory (CFT)
\cite{aha}. There are  many works on AdS black holes on this
subject \cite{horo,car1,moss,wang,mmz}. However, most of  the
works on QNMs of black holes in four and higher dimensions  are
numerical except for a few cases~\cite{aros,birs,ort1,ort2}. On
the other hand, QNMs of the  BTZ black hole~\cite{banados} has
been studied with exact results in AdS$_3$
space~\cite{bir1,bir2,car2,abd}. In this case, the black hole
corresponds to a thermal state in the CFT$_2$, and the decay of
the test perturbation in the black hole spacetime corresponds to
the decay of the perturbed state in the CFT$_2$. Actually, the
imaginary part of the frequency $\omega_i$, which determines how
damped the mode, is a measure of the characteristic time
$\tau=1/\omega_i$ of approach to thermal equilibrium through the
AdS$_3$/CFT$_2$ correspondence.

Recently, the purely imaginary quasinormal frequencies of a
massless scalar perturbation on the 3D charged-dilaton black holes
(CDBHs) were found  using the scattering picture for
$E>V_0$~\cite{fer1,fer2,fer08} with the energy $E=\omega^2$ and
the height of the potential-step $V_0$. As far as we know, three
cases \cite{car1,fer1,fer2,fer08} for these quasinormal
frequencies including the electromagnetic and odd-gravitational
perturbations on the 4D Schwarzschild-AdS (SAdS) black holes are
known to date. It seems that as long as the modes are decaying, it
does not matter whether they are oscillating or not. However, the
purely imaginary quasinormal frequencies represent a special set
of modes which are purely damped~\cite{mmz}. These are not regular
QNMs because the real part of frequencies vanishes, eliminating
the oscillatory behavior of the perturbations which is
characteristic of QNMs. In other words, the purely imaginary
quasinormal frequencies should be distinguished from the regular
QNMs.

Hence, it is very important to explore the connection between the
purely imaginary quasinormal frequencies and the potential. Most
of potentials surrounding the black holes  is the barrier-type
localized at $r_*=0$ in asymptotically flat spacetimes, whereas
most of  black holes has monotonically increasing potentials in
asymptotically AdS spacetimes. However, the potentials surrounding
the CDBH are the step-type, like the potentials surrounding  the
4D SAdS black holes for the electromagnetic and odd-gravitational
perturbations~\cite{car1}.

In this work,  we explore the connection between the quasinormal
frequencies and the shape of potentials using  the scattering
picture for $E>V_0$. It turns out that the potential-step type
provides the purely imaginary quasinormal frequencies, while the
potential-barrier type gives the complex QNMs. On the other hand,
for $E<V_0$ which is the case of de Sitter space~\cite{Mkim}, we
could not find any QNM.

\section{Potentials surrounding black holes}

\subsection{Potential-step}

\begin{figure}[t!]
   \centering
   \includegraphics{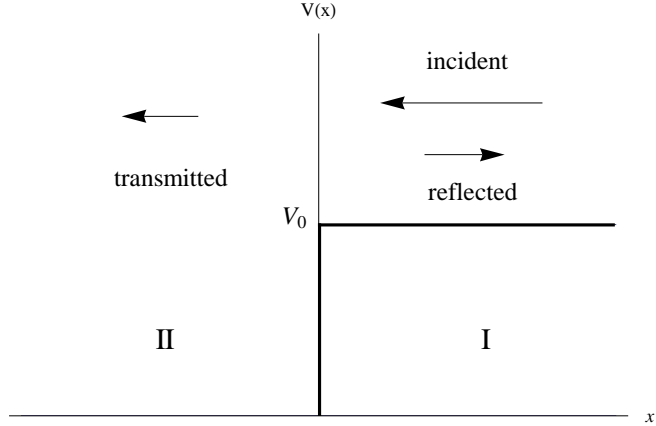}
\caption{Potential-step: the incident wave comes from the right
with $E\ge V_0$} \label{fig1}
\end{figure}

\begin{figure}[t!]
   \centering
   \includegraphics{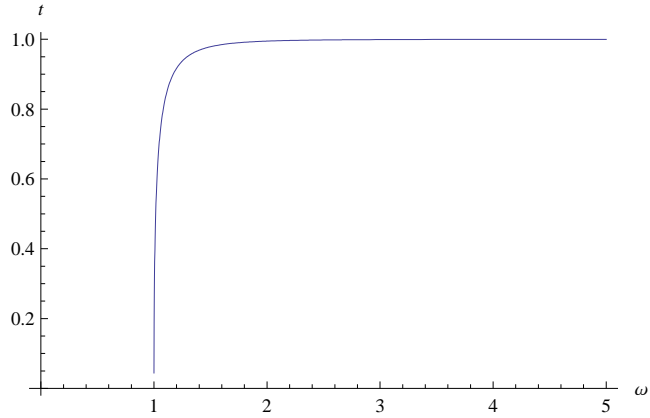}
\caption{Transmission  coefficient $t$ for the potential-step: for
$m=1/2$, $\hbar=1$, and $V_0=1$. Note that $\omega\ge 1$. }
\label{fig2}
\end{figure}

Let us consider a toy model of wave propagation under a
potential-step with a height $V_0(<E)$ as shown in Fig.~\ref{fig1}.
By solving the Schr\"odinger equation when a wave propagates from
right to left, one can easily obtain the incident, reflected, and
transmitted flux as
\begin{eqnarray}
{\cal F}_{in}&=& -2\sqrt{\omega^2-V_0}, \nonumber\\
{\cal F}_{re}&=& 2\sqrt{\omega^2-V_0}\left|R\right|^2, \nonumber\\
{\cal F}_{tr}&=&-2\omega \left|T\right|^2,
\end{eqnarray}
respectively,  with $\omega^2=E$, $\hbar=1$ and  $m=1/2$. Here the
$R$ reflection and $T$ transmission amplitudes are given by
\begin{equation}
R=\frac{\sqrt{\omega^2-V_0}-\omega}{\sqrt{\omega^2-V_0}+\omega},~~
T=\frac{2\sqrt{\omega^2-V_0}}{\sqrt{\omega^2-V_0}+\omega}.
\end{equation}
Then, the $r$ reflection and $t$ transmission coefficients are
\begin{eqnarray}
r=\left|\frac{{\cal F}_{re}}{{\cal F}_{in}}\right|&=&
   \frac{(\sqrt{\omega^2-V_0}-\omega)^2}{(\sqrt{\omega^2-V_0}+\omega)^2}, \nonumber\\
t=\left|\frac{{\cal F}_{tr}}{{\cal F}_{in}}\right|
                                        &=&\frac{4\omega\sqrt{\omega^2-V_0}}{(\sqrt{\omega^2-V_0}+\omega)^2},
\end{eqnarray}
respectively, satisfying with $r+t=1$, which means that the flux
are conserved. When the  energy of the incident wave increases,
the transmission coefficient approaches unity, as shown in
Fig.~\ref{fig2}. On the other hand, as $\omega$ approaches
$\sqrt{V_0}$, one has no transmission and total reflection.

\subsection{3D charged-dilaton black holes (CDBH)}
The starting action with the dilaton field $\phi$ is given by
\begin{equation}
S=\int d^3x
\sqrt{-g}\Big[R+2e^{4\phi}\Lambda-4(\nabla\phi)^2-e^{-4\phi}F_{\mu\nu}F^{\mu\nu}
\Big]
\end{equation}
with the cosmological constant $\Lambda>0$ for anti-de Sitter
spacetimes. This action is conformally related to the low-energy string action
in three dimensions. The solution of the CDBH~\cite{CM} is given by
\begin{equation}
ds^2=-f(r)dt^2+\frac{4r^2 dr^2}{\gamma^4 f(r)}+r^2
d^2\theta;~~\phi=\frac{1}{4} \ln
\Big[\frac{r}{\gamma^2}\Big];~~F_{rt}=\frac{Q}{r^2},
\end{equation}
where the metric function $f(r)=-2Mr+8\Lambda r^2+8Q^2$ and an
integration constant $\gamma$ with dimension ${\rm
[L]}^{\frac{1}{2}}$ is necessary to have correct dimensions. Then,
two horizons can be obtained from the condition of $f(r)=0$ as
\begin{eqnarray}
r_\pm=\frac{M\pm\sqrt{M^2-64Q^2\Lambda}}{8\Lambda},
\end{eqnarray}
where $r_+(r_-)$ is the outer (inner) horizon for $M\ge
8Q\sqrt{\Lambda}$. The degenerate horizon of $r_e=r_+=r_-$ appears
when $M=8Q\sqrt{\Lambda}$. The thermodynamics of CDBH will be
discussed in Appendix.

In the background of the CDBH, we introduce a scalar perturbation
$\Phi$, which satisfies the Klein-Gordon equation
\begin{equation}
\Big({\nabla}^2-\mu^2\Big)\Phi=0.
\end{equation}
Hereafter we consider the massless case of $\mu=0$ only because
the massive case gives rise to some difficulties with analytic
calculations. Using the cylindrical symmetry of the background, let us
parameterize $\Phi$ as
\begin{equation}
\Phi(t,r,\theta)=e^{i\omega t}e^{im\theta}\xi(r).
\end{equation} Then, the above equation leads to the radial
equation
\begin{equation} \label{radialeq}
\frac{d}{dr}\left(\frac{\gamma^4f(r)}{2}\frac{d}{dr}\xi(r)\right)
+2r^2\left(\frac{\omega^2}{f(r)}-\frac{m^2}{r^2}\right)\xi(r)=0.
\end{equation}
Redefining the radial function $\xi(r)$ as
\begin{equation}
\xi(r)\equiv\frac{R(r)}{\sqrt{r}}
\end{equation}
and introducing the tortoise coordinate defined by $dr_*=2rdr/\gamma^2f(r)$,
the radial equation of the massless scalar perturbation becomes
the Schr\"odinger equation with the energy $E=\omega^2$
\begin{equation}
\label{scheq} \left(\frac{d^2}{dr^2_*}+E-V(r)\right) R(r)=0.
\end{equation}
\begin{figure}[t!]
   \centering
   \includegraphics{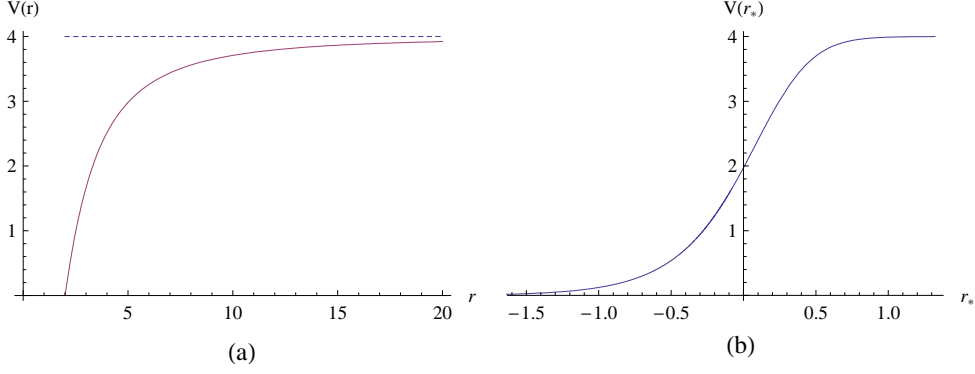}
\caption{(a) The potential of the CDBH for $M=10$, $Q=1$,
$\Lambda=1$, $m=0$, and $\gamma=1$. (b) The corresponding
potential depicted in the tortoise coordinate $r_*$. }
\label{fig3}
\end{figure}
\begin{figure}[t!]
   \centering
   \includegraphics{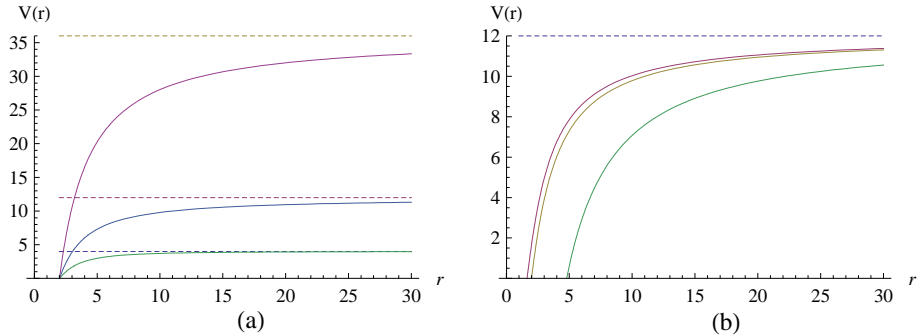}
\caption{(a) The potentials for different angular momentum number
$m$ with $M=10$, $Q=1$, $\Lambda=1$, and $\gamma=1$. The curves
are for $m=2$, $m=1$, and $m=0$ from top to bottom, and start from
the same $r_+=2$. (b) The potentials for different masses $M$ with
$Q=1$, $\Lambda=1$, $m=1$, and $\gamma=1$. The curves are for
$M=9$ ($r_+=1.6$), $M=10$ ($r_+=2.0$), and $M=20$ ($r_+=4.8$) from
top to bottom. All curves approach the maximum $V_0$ given by
Eq.~(\ref{minV}) as $r$ increases.} \label{fig4}
\end{figure}
Here, the potential $V(r)$ is given by
\begin{eqnarray} \label{potential}
V(r)&=& \gamma^4\left(-\frac{12Q^4}{r^4}+\frac{4M Q^2}{r^3}
      -\frac{M^2}{4r^2}-\frac{8\Lambda  Q^2}{r^2}+ 4\Lambda^2\right) \nonumber\\
      &+&\frac{m^2}{r^2}\left(-2M r+8\Lambda r^2+8 Q^2\right).
\end{eqnarray}
This potential is depicted in Fig.~\ref{fig3}(a). The potential
is monotonically increasing  outside the event horizon and
approaching a constant value determined by the height of the
potential step
\begin{equation}
\label{minV} V_0(m)=8m^2\Lambda+4\gamma^4\Lambda^2
\end{equation}
for a given  $m$. On the other hand, for the CDBH case the
tortoise coordinate takes the form
\begin{equation}
\label{tortoise1} r_*=\frac{1}{4\gamma^2\Lambda(r_+-r_-)}
[r_+\ln(r-r_+)-r_-\ln(r-r_-)].
\end{equation}
Then, when $r\rightarrow r_+$, the tortoise coordinate $r_*$
approaches $-\infty$, while when $r\rightarrow\infty$, the
tortoise coordinate $r_*$ goes to $\infty$.

Now, in order to see the shape of the potential (\ref{potential})
expressed in terms of the tortoise coordinate $r_*$, let us draw the
effective potential $V(r_*)$ in Fig.~\ref{fig3}(b) numerically. We
observe from this graph that the effective potential is different
from the Schwarzschild-type one, which has a shape of the barrier.
For the CDBH, the shape of the potential is the step-type
localized near $r_*=0$, while for the Schwarzschild black hole, it
looks like  the barrier.  Also we find similar graphs from
Fig.~\ref{fig4}  for different $m$ and $M$.

From now on we separate the case of $E > V_0$ from the case of
$E < V_0$ because the former has a completely different scattering
picture from the latter. In order to compute the greybody factor
and QNMs, we have to solve the radial equation (\ref{radialeq})
completely by introducing a new coordinate
\begin{equation}
z=\frac{r-r_+}{r-r_-},
\end{equation}
which covers a compact region of $0\le z\le 1$, corresponding to
$-\infty \le r_* \le \infty~(r_+ \le r \le \infty)$. Then,
Eq.~(\ref{radialeq}) leads to
\begin{equation}
\label{primaryeq}
z(1-z)R''(z)+(1-z)R'(z)+\frac{1}{\gamma^4}P(z)R(z)=0,
\end{equation}
where
\begin{equation}
P(z)=A+\frac{B}{z}+\frac{C}{z-1}
\end{equation}
with
\begin{equation}
A=-\frac{r_-^2\omega^2}{16\Lambda^2(r_+-r_-)^2},~~
B=\frac{r_+^2\omega^2}{16\Lambda^2(r_+-r_-)^2},~~
C=\frac{\omega^2-8m^2\Lambda}{16\Lambda^2}.
\end{equation}
With an ansatz of $R(z)=z^\alpha(1-z)^\beta F(z)$, the radial
equation (\ref{primaryeq}) is written by
\begin{eqnarray}
\label{hyeq}
&& z(1-z)F''(z)+[1+2\alpha-(1+2\alpha+2\beta)z]F'(z) \nonumber\\
&&+\left(\frac{A}{\gamma^4}-(\alpha+\beta)^2+\frac{B/\gamma^4+\alpha^2}{z}
+ \frac{C/\gamma^4+\beta^2-\beta}{1-z}\right)F(z)=0.
\end{eqnarray}
Comparing this with the hypergeometric equation \cite{Abramowitz}
given by
\begin{equation}
\label{diffeq2}
z(1-z)\frac{d^2F}{dz^2}+[c-(1+a+b)]\frac{dF}{dz}-abF=0,
\end{equation}
we have
\begin{eqnarray}
\label{coeff1} && c=1+2\alpha,\\
\label{coeff2} && a+b=2\alpha+2\beta, \\
\label{coeff3} && ab=(\alpha+\beta)^2-A/\gamma^4,\\
\label{coeff4} && B/\gamma^4+\alpha^2=0,\\
\label{coeff5} &&C/\gamma^4+\beta^2-\beta=0.
\end{eqnarray}
By solving Eqs. (\ref{coeff2}) and (\ref{coeff3}), we have
\begin{equation}
a\equiv
  a_{\pm}=\alpha+\beta\pm\frac{\sqrt{A}}{\gamma^2},~~~
b\equiv
   b_{\mp}=\alpha+\beta\mp\frac{\sqrt{A}}{\gamma^2},
\end{equation}
respectively. One finds from Eqs. (\ref{coeff4})and (\ref{coeff5}),
\begin{equation}
\label{alpha}
  \alpha\equiv\alpha_{\pm}=\pm\frac{ir_+\omega}{4\gamma^2\Lambda(r_+-r_-)},
\end{equation}
\begin{equation}
\label{beta} \beta\equiv\beta_{\pm}=\frac{1}{2}\Bigg[1\pm
i\sqrt{\frac{\omega^2-V_0}{4\gamma^4\Lambda^2}}\Bigg],
\end{equation}
respectively. Hereafter we will use the upper signs of $\alpha_+$ and $\beta_+$
without loss of generality. For $\omega^2<V_0$, we have a
completely different case of  $\beta$=real.
This will be discussed in Sec. 4 separately. Since the
hypergeometric  equation (\ref{hyeq}) has regular singular points
at $z=0$, $z=1$ and none of $a$, $b$, and $c$ are integer, there
exists the general solution in the neighborhood of the points
\cite{Abramowitz} such as
$C_1F(a,b,c;z)+C_2(1-z)^{1-c}F(a-c+1,b-c+1,2-c;z)$ with two
unknown constants $C_1$ and $C_2$.

Now, in order to find the scattering picture of a massless scalar
on the CDBH, we need to specify boundary conditions: near horizon
and at infinity. We choose the condition that the wave is purely
ingoing $(\longleftarrow)$ near the horizon. Then, we find the
ingoing mode (transmitted mode) near $z=0$ by choosing $C_2=0$ as
\begin{equation}
\label{nearextS}
 e^{i\omega t}R(r_*\rightarrow -\infty)=\frac{C_1}{|r_+-r_-|^{\alpha}}
 e^{i\omega(t+r_*)}.
\end{equation}
On the other hand, the asymptotic solution matched with
Eq.~(\ref{nearextS}) is
\begin{eqnarray}
\label{exactSasym}
 e^{i\omega t}R(r_*\rightarrow\infty)&=&
  D_1~e^{i\omega\left(t - r_*\sqrt{1-\frac{V_0}{\omega^2}}\right)-2\gamma^2\Lambda r_*} \nonumber\\
  &+&D_2~e^{i\omega\left(t + r_*\sqrt{1-\frac{V_0}{\omega^2}}\right)-2\gamma^2\Lambda r_*},
\end{eqnarray}
where $D_1$, $D_2$ are defined by
\begin{eqnarray}
  &&D_1= C_1\left(\frac{r_+-r_-}{r_+}\right)^{\beta}\frac{\Gamma(1+2\alpha)\Gamma(1-2\beta)}
      {\Gamma(1+\alpha-\beta-\sqrt{A}/\gamma^2)\Gamma(1+\alpha-\beta+\sqrt{A}/\gamma^2)}, \nonumber\\
  &&D_2= C_1\left(\frac{r_+-r_-}{r_+}\right)^{1-\beta}\frac{\Gamma(1+2\alpha)\Gamma(2\beta-1)}
      {\Gamma(\alpha+\beta+\sqrt{A}/\gamma^2)\Gamma(\alpha+\beta-\sqrt{A}/\gamma^2)},
\end{eqnarray}
respectively. Note that the first $D_1$ term in
Eq.~(\ref{exactSasym}) is an outgoing mode ($\longrightarrow$) and
the second $D_2$ term is an ingoing mode ($\longleftarrow$) at
infinity.
\begin{figure}[t!]
   \centering
   \includegraphics{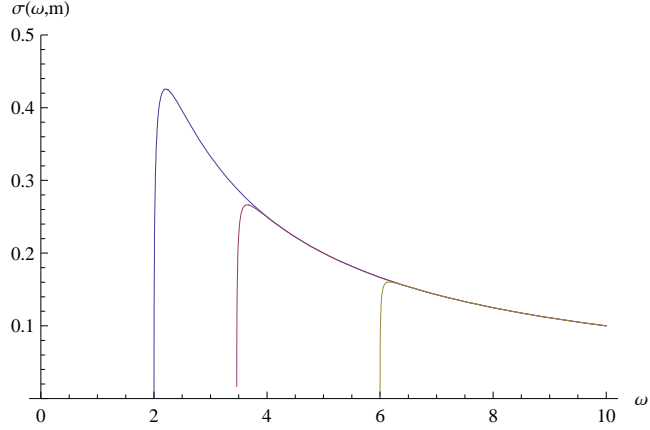}
\caption{Plots of the partial wave absorption cross section for
$M=10$, $Q=1$, $\Lambda$, and $\gamma=1$: the curves are depicted
for $m=0$, $m=1$, and $m=2$ from left to right.} \label{fig5}
\end{figure}
\begin{figure}[t!]
   \centering
   \includegraphics{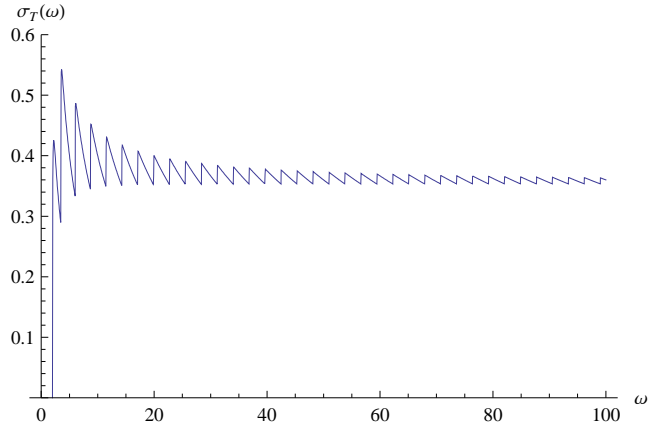}
\caption{Plot of total absorption cross section for $M=10$, $Q=1$,
$\Lambda=1$, and $\gamma=1$. It is obtained by summing
$\sigma(\omega,m)$ from $m=0$ to $m=35$.} \label{fig6}
\end{figure}

Using these solutions, one can calculate the partial wave
absorption cross section as
\begin{eqnarray}
\sigma(\omega,m)&=&\frac{1}{\omega}
  \left|\frac{{\cal F}_{tr}(r\rightarrow r_+)}{{\cal F}_{in}(r\rightarrow\infty)} \right|
  =\frac{8\pi r_+|C_1|^2/\gamma^2}
     {16\pi|D_2|^2(r_+-r_-)\Lambda\sqrt{\frac{V_0-\omega^2}{4\gamma^4\Lambda^2}}}\\
 &=&\frac{\gamma^2\sinh\left(\frac{\pi \omega r_+}{2\gamma^4\Lambda^2}\right)
      \sinh\left(\pi\sqrt{\frac{\omega^2-V_0}{4\gamma^4\Lambda^2}}\right)}
      {\omega \cosh\left(\frac{\pi}{2}\sqrt{\frac{\omega^2-V_0}
      {4\gamma^4\Lambda^2}}+\frac{\pi\omega}{4\gamma^2\Lambda}\right)
      \cosh\left(\frac{\pi}{2}\sqrt{\frac{\omega^2-V_0}{4\gamma^4\Lambda^2}}
      +\frac{\pi\omega(r_++r_-)}{4\gamma^2\Lambda(r_+-r_-)}\right)} \nonumber,
\end{eqnarray}
which is depicted for $m=0,1,2$ in Fig.~\ref{fig5}. We confirm that
$\sigma(\omega=\sqrt{V_0},m)=0$, which shows no transmission
clearly. This contrasts to the case of the Schwarzschild black
hole, which shows the universal behavior of $\sigma^{SBH}
(\omega\to 0) \to {\cal A}_H$ with the area of horizon ${\cal
A}_H$ for a massless scalar propagation.

On the other hand, the total absorption cross section given by
\begin{equation}
\sigma_{T}(\omega)=\sum_{m}\sigma(\omega,m)
\end{equation}
is depicted in Fig.~\ref{fig6}. Note that the  maximum of each
partial absorption cross section makes a local maximum in the
total absorption cross section. As $\omega$ increases,
$\sigma_{T}$ approaches a constant value, $2\pi/17=0.37$. This
oscillatory behavior in the total absorption cross section
reflects the wiggly characteristic of the potential-steps
surrounding the CDBH.
Note that similar oscillatory behavior also appear
in the total absorption cross sections of the Schwarzschild black hole~\cite{Sanchez,dkp}
and the $D3$-brane~\cite{Cvetic}.

In the case of the potential-barriers surrounding the
Schwarzschild black hole, the total cross section is given
by~\cite{Sanchez}
\begin{equation}
\sigma_{T}^{SBH}(\omega)=\frac{27\pi}{4}-\frac{\sqrt{2}\sin(\sqrt{2\pi^3}\omega)}{\omega}
\end{equation}
which indicates the limiting value of $27\pi/4=21.2$ for large
$\omega$. Hence, the high energy limit ($\omega \gg \sqrt{V_0}$)
of the total cross section of the CDBHs is also similar to the
case of the Schwarzschild black hole, even though the latter shows
the potential-barriers with different heights for different
angular momentum number $\ell$. However, for the CDBH case the low
energy limit of $\omega \to \sqrt{V_0}$ is quite different from
the case of the Schwarzschild black hole.

\subsection{The Schwarzschild-AdS black holes (SAdS) }
\begin{figure}[t!]
   \centering
   \includegraphics{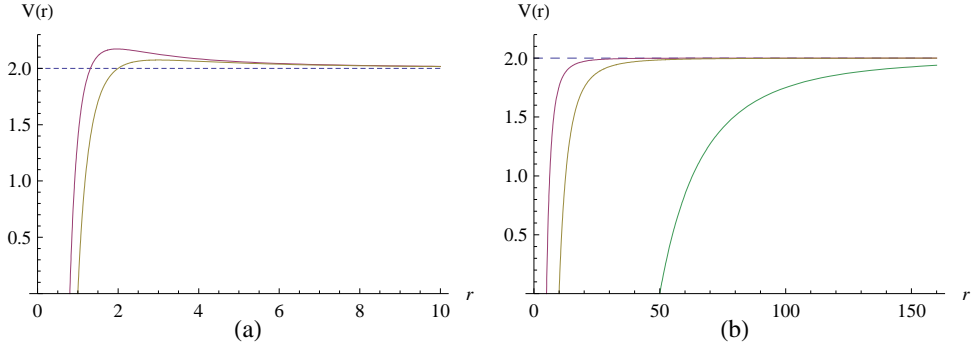}
\caption{Two different types of potentials for the SAdS black
holes with $R=1$, $l=1$. (a) The barrier-type for small black
holes with $r_+<5$: Two curves from left to right are for
$M=0.656(r_+=0.8)$ and $M=1(r_+=1)$, which have a bump rising
above the asymptotic value of the potential $V_{0SAdS}=2$. (b) The
step-type for large black holes with $r_+\ge5$: Three curves under
$V_{0SAdS}=2$ are for $M=65$ ($r_+=5$), $M=505$ ($r_+=10$), and
$M=62505$ ($r_+=50$) from left to right.}
 \label{fig7}
\end{figure}

It was reported that the electromagnetic perturbations on the SAdS
spacetime shows the purely imaginary QNM frequencies~\cite{car1}.
Let us study the propagation of an electromagnetic field in the
SAdS with the line element
\begin{equation}
ds^2=-f(r)dt^2+\frac{dr^2}{f(r)}+r^2d^2\Omega,
\end{equation}
where $f(r)=1-2M/r+r^2/R^2$ and $R$ is the AdS radius.
An electromagnetic wave on this background is described by Maxwell
equation, which turns out to be  the  Schr\"odinger-like equation of the
type (\ref{scheq}) for the radial part. The general form is given
by
\begin{equation}
\frac{d^2\psi}{dr_*^2}+\Big[\omega^2 -V_{SAdS}(r) \Big]\psi=0,
\end{equation}
where the potential is given by
\begin{equation}
V_{\rm
SAdS}(r)=f(r)\Big[\frac{l(l+1)}{r^2}+\frac{\beta}{r^3}\Big].
\end{equation}
Here $\beta$ is $-6M,0$, and $2M$ for odd-gravitational, electromagnetic
and scalar perturbations, respectively. For the $\beta=0$ case,
its shape is depicted in Fig.~\ref{fig7}.
As shown in this figure, the potential for a large SAdS black hole
approaches a constant value given by $V_{0SAdS}=l(l+1)/R^2$ of a
step-type, while in the Schwarzschild limit
($R\rightarrow\infty$), its asymptotic tail goes to zero, making a
barrier-type potential. Note here that  there exists a local bump
near the horizon, depending the mass of the black hole. For the
small black hole with $r_+<5$, the potential has the  maximum
value near the horizon, while for the large black hole with
$r_+>5$, the maximum disappears and its shape becomes  the
potential-step. This is similar to the massless scalar
perturbation on the CDBH in the previous section.

\section{Quasinormal modes}
In general, the QNMs can be decomposed into real and imaginary
parts as follows:
\begin{equation}
\omega=\omega_r-i\omega_i.
\end{equation}
In this section, we provides two cases of purely damped modes with
$\omega_r=0$. We explore the connection between the purely
imaginary QNMs and the potential-steps surrounding the black
holes.
\subsection{QNMs in the SAdS}

It is by now well-known that the electromagnetic perturbation on the SAdS
shows the different feature that its QNMs take
purely imaginary values for a large black hole \cite{car1}.
Here we quote their results in Table~\ref{table1} that shows the
lowest QNMs for the $l=1$ electromagnetic perturbation on the
SAdS spacetime.

\begin{table}
\centering
\begin{tabular}{|c|c|c|}
 \hline\hline
 \multicolumn{3}{|c|}{Numerical Results} \\
 \cline{1-3}
\hline\hline
  $r_+$ & $-\omega_i$ & $\omega_r$ \\
 \cline{1-3}
 0.8 & 1.287 & 2.175 \\ \hline
 1 & 1.699 & 2.163 \\ \hline
 5 & 8.795 & $\sim$ 0 \\ \hline
 10 & 15.506 & $\sim$ 0 \\ \hline
 50 & 75.096 & $\sim$ 0 \\ \hline
 100 & 150.048 & $\sim$ 0 \\ \hline
\end{tabular} \caption{In the case of $l=1$, the QNMs frequencies of electromagnetic perturbation
for selected values of $r_+$ on the SAdS.} \label{table1}
\end{table}

We observe that starting from $r_+=5$, they have imaginary values.
Note that for the small black holes with $r_+ <5$, the shape of
potential is the barrier-type. See Fig.~\ref{fig7}(a). On the
other hand, for the large black holes with $r_+ \ge 5$, as shown
in Fig.~\ref{fig7}(b), the corresponding potentials are the
step-type. In order to have complex QNMs, the potential should
have  a bump (local maximum) somewhere. The presence of bump near
the horizon  explains clearly why the QNMs of the small
Schwarzschild-AdS black hole are complex~\cite{car1}.

\subsection{QNMs in the CDBH}

The QNMs are the solutions of the wave equation characterized by
purely ingoing waves  near the horizon. In addition, one has to
impose the boundary condition at infinity. Thus, one needs to know
the asymptotic geometry of the spacetime under study, which is
 provided by the  potential. As for the
CDBH, the author~\cite{fer2,fer08} took the boundary condition for
the asymptotic region as the AdS spacetime. However, as seen from
the potential in Fig.~\ref{fig3}(a) and \ref{fig3}(b), the energy
of a perturbed scalar  should be higher than the maximum effective
potential $V_0$ for $E>V_0$. On the other hand, the propagating wave
could not be developed for $E<V_0$. Thus, a  boundary condition for the CDBH
seems to be  the one in asymptotically flat spacetime: purely
outgoing waves at infinity \cite{horo}. Precisely, as is shown in
Eq. (\ref{exactSasym}), it is a combination of asymptotically flat
spacetimes (plane waves) and AdS spacetimes ($e^{-2
\gamma^2\Lambda r_*}$-normalizable mode).

Considering the asymptotic solution Eq.~(\ref{exactSasym}),
keeping the first term $(D_1\not=0,D_2=0$)  corresponds to the
outgoing mode only. Equivalently, considering the scattering
resonances at the pole of the transmission amplitude ($D_2 \to 0$)
leads to the condition for QNMs for the massless scalar
on the CDBH as
\begin{equation}
\alpha+\beta \pm \frac{\sqrt{A}}{\gamma^2}=-n
\end{equation}
with the positive integer $n$.
From this condition, we obtain the quasinormal frequencies
\begin{eqnarray}
\label{qnm3d}
\omega_n&=&-i\frac{\gamma^2\Lambda(r_+-r_-)}{2r_+r_-}
\Big[(2n+1)(r_++r_-)\Big.\nonumber\\
&+&
\Big.\sqrt{(2n+1)^2(r_++r_-)^2-16n(n+1)r_+r_-+\frac{8m^2r_+r_-}{\gamma^4\Lambda}}\Big]\\
&=&-i\frac{\pi T_H}{2}\left[(2n+1)(1+\frac{r_+}{r_-})\right.\nonumber\\
&+&
\left.\sqrt{(2n+1)^2(1+\frac{r_+}{r_-})^2-16n(n+1)r_++\frac{8m^2r_+}{\gamma^4\Lambda}}\right],
\end{eqnarray}
where the Hawking temperature is given by
$T_H=\gamma^2\Lambda(r_+-r_-)/\pi r_+$. We prove that $\omega_n$
is purely imaginary by numerically showing that the quantity
inside the square root is always positive for any CDBH.

\section{$E<V_0$ case in the CDBH}
This case corresponds to the purely real $\beta$ case and in turn
its flux becomes zero. Let us explain this situation by
introducing a wave propagation from left to right under a
potential-step with height $V_0(> E=\omega^2)$ in Fig. 1.
 Since the density of incident wave is
unity, its flux (${\cal F}_{II in}$) is equal to $2\omega$. The
reflected flux ${\cal F}_{ II re}$ is
 given by $- 2\omega$ and thus there is no
net flux in region II: ${\cal{F}}_{II in}+{\cal{F}}_{ II re}=0$.
According to the flux conservation, we expect that there is no
flux in region I. As a check, one finds that ${\cal F}_{
I}=\frac{1}{i} [\Psi^{*}_{\rm I}(\Psi_{\rm I})'- \Psi_{\rm I}(\Psi
^{*}_{\rm I})']=0$. Even though the probability density of finding
a particle between $x$ and $x+dx(x>0)$ is not zero, its flux is
zero. This means that the quantum mechanical picture reduces to
the classical picture of the total reflection. A plane wave moving
under a potential-step with the height $V_0(>E$) corresponds to a toy
model for the total reflection with $|R|^2=1$ and
$~R=e^{-2i\theta}$ (non-zero phase). A similar situation occurs in
a scalar wave propagating under de Sitter space~\cite{Mkim}.

In this case we have no absorption cross section of
$\sigma(\omega,m)={\cal F}_I/\omega {\cal F}_{II re}=0$  because
of zero flux ${\cal F}_I=0$. Furthermore, QNMs are not defined
since we could not make ${\cal F}_{II in}=0$ without ${\cal F}_{II
re}=0$ .   This result is consistent with the picture of stable
event horizon because the presence of quasinormal frequencies
implies that  the massless scalar wave is loosing its energy
continuously into the event horizon.  This means that the event
horizon  is usually stable  and it is always in thermal
equilibrium with the scalar perturbation. More explicitly, the
event horizon not only absorbs scalar waves  but also emits those
previously absorbed by itself at the same rate.
As a result, the  case of $E<V_0$ corresponds to a
forbidden region for scattering (absorption cross section) and
scattering resonances (QNMs).

\section{Discussions}
We clarify the purely imaginary quasinormal frequencies of a
massless scalar perturbation on the 3D charged-dilaton black
holes(CDBH) using the scattering (resonance) picture for $E>V_0$
with the energy $E=\omega^2$ and the height $V_0$ of the potential
step. In fact, the purely imaginary quasinormal frequencies
represent a special set of modes which are purely
damped~\cite{mmz}. These are not regular QNMs because the real
part of frequencies vanishes, eliminating the oscillatory behavior
of the perturbations which is characteristic of QNMs.

This CDBH case is quite interesting because we have found the potential-steps
outside the event horizon,  similar to the case of the
electromagnetic and odd-gravitational perturbations on the large
Schwarzschild-AdS black holes. We believe that  the potential
surrounding the black hole provides all information to a
perturbation field propagating on the black hole background.

Moreover, we have  shown  that the potential-step type provides
the purely imaginary quasinormal frequencies, while the
potential-barrier type gives the complex quasinormal frequencies.
For $E<V_0$, which is the case of de Sitter space, we could not
find any QNMs.

Finally, we discuss the connection between thermodynamics of black
hole and QNMs. As is shown in Appendix, we have shown that
thermodynamics of the CDBH has a new feature comparing with the
non-rotating BTZ black hole, which provides the complex QNMs. This
includes a constant value of temperature  and different behaviors
of mass and heat capacity for the large CDBH. Even though
thermodynamics of the CDBH is different from that of the
non-rotating BTZ black hole, it is not clear why the QNMs of the
CDBH are purely imaginary whereas the QNMs of the non-rotating BTZ
black hole are complex. However, comparing their potentials leads
to the fact that the potentials of the CDBH are step-type, while
those of the non-rotating BTZ black hole are monotonically
increasing functions. The latter are infinite at
infinity~\cite{Mlee}.

Consequently, we have shown that the purely imaginary quasinormal
frequencies of a massless scalar is determined by its
potential-steps surrounding the charged-dilaton black hole.

\section*{Appendix: Thermodynamics of the CDBH}
\begin{figure}[t!]
   \centering
   \includegraphics{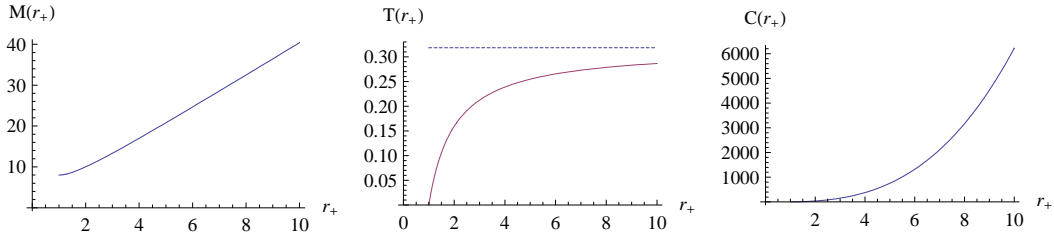}
\caption{Graphs of mass, temperature, and specific heat of the
CDBH for $Q=1$, $\Lambda=1$, and $\gamma=1$, respectively.}
\label{fig9}
\end{figure}
In this Appendix, we consider the thermodynamic quantities of the CDBH.

The mass of the CDBH is given by
\begin{equation}
M(r_+,Q,\Lambda) =
   4\Lambda r_+ \left( 1 + \frac{Q^2}{\Lambda r_+^2} \right).
\end{equation}
For large black hole, it  is  given by $M \propto 4\Lambda r_+$.

The Hawking temperature is
\begin{equation}
T(r_+,Q,\Lambda)=
   \frac{\gamma^2\Lambda}{\pi}\left(1 -\frac{Q^2}{\Lambda
   r^2_+}\right).
\end{equation}
For the large black hole, the temperature goes to a constant value of
$\gamma^2\Lambda/\pi$, which is the temperature of the
uncharged-dilaton black hole. This is  a new feature of the CDBH.

The specific heat is
\begin{equation}
C(r_+,Q,\Lambda)=\frac{2\pi\Lambda r_+^3}{Q^2}
   \left(1-\frac{Q^2}{\Lambda r_+^2}\right).
\end{equation}
The heat capacity has a single positive phase like the
non-rotating BTZ black hole but there exists a difference in their
forms. Note that the thermodynamic quantities of the non-rotating BTZ black
hole are given  by $M_{NBTZ}=\Lambda r_+^2,~T_{NBTZ}=\Lambda
r_+/2\pi,~C_{NBTZ}=S_{NBTZ}=4\pi r_+$ with $\Lambda=1/l^2$.
The extremal mass $M_e$ is given by
$8Q\sqrt{\Lambda}$ at the degenerate horizon
$r_e=Q/\sqrt{\Lambda}$, where the temperature and specific heat
vanish.

Finally, the Bekenstein-Hawking entropy is given by the area-law
as
\begin{equation}
S_{BH}= \frac{2\pi r_+}{4G}=4\pi r_+
\end{equation}
with $G=1/8$. Then, we can easily check that the first-law of
thermodynamics is satisfied as
\begin{equation}
\gamma^2 dM = TdS_{BH}.
\end{equation}

\section*{Acknowledgment}
Two of us (Y. S. Myung and Y.-J. Park) were supported by the
Science Research Center Program of the Korea Science and
Engineering Foundation through the Center for Quantum Spacetime of
Sogang University with grant number R11-2005-021.
Y.-W. Kim was supported by the Korea Research Foundation Grant
funded by Korea Government (MOEHRD): KRF-2007-359-C00007.

\end{document}